\documentstyle[12pt]{article}
\begin{document}

\setcounter{footnote}{0}
\setcounter{equation}{0}
\setcounter{figure}{0}
\setcounter{table}{0}
\vspace*{5mm}

\newcommand{\alt}{\mathbin{\lower 3pt\hbox
   {$\rlap{\raise 5pt\hbox{$\char'074$}}\mathchar"7218$}}}
\newcommand{\agt}{\mathbin{\lower 3pt\hbox
   {$\rlap{\raise 5pt\hbox{$\char'076$}}\mathchar"7218$}}}

\begin{center}

{\Large\bf Numerical results for the Anderson transition. Comment
} 

\vspace{4mm}
I. M. Suslov \\
P.L.Kapitza Institute for Physical problems \\
117334, Moscow, Russia  \\
E-mail: suslov@kapitza.ras.ru
\\ \vspace{6mm}
\end{center}

\begin{center}
\begin{minipage}{135mm}
{\large\bf Abstract } \\
Answer to cond-mat/0106005, cond-mat/0106006 and additional notes 
are given concerning my previous comment (cond-mat/0105325).

\end{minipage}
\end{center}

\vspace{15mm}

Recently I have discussed the numerical results for the Anderson transition
\cite{1}. Objections \cite{2} and \cite{2a} have appeared after that.
We answer here to the arguments of \cite{2,2a} and give the additional
comments.

\vspace{3mm}

\vspace{5mm}

1. I did not state in \cite{1} that $\nu =1$ (in fact, my most probable 
value was 1.25).  I tried to show that a problem exists: numerical results 
contradict to all other information on the critical exponents. It is not my 
personal problem, and it is not my personal affair to solve it.  Indeed, a 
value $\nu \approx 1$ is desirable from viewpoint of this other information 
and I tried to understand, is not it possible to agree it  with the raw 
numerical data by the change of interpretation and by more realistic 
estimation of errors.  I found it possible for the data that I considered as 
the best. May be I was a little wrong in the latter estimation.

Authors of \cite{2} argue that I used the systems of too 
small size ($28^3$), "while currently sizes $\sim 50^3$ (for ELS) and 
$18^2\times 10^8$ (for TM) are standard"\,\footnote{\,Abbreviations of
\cite{2}: TM and ELS --- transfer matrix and energy level statistics
methods, FSS --- finite size scaling.}. A size $18^2\times 10^8$
should kick a man out his feet, while authors of \cite{2} know perfectly
well that only the minimal length scale ($L=18$) is relevant. In fact, such 
systems {\it are small}.\,\footnote{\, I can understand such trick in the 
text, where authors of \cite{2} are in the course of discussion, but the same 
thing is present  in the figure capture:  "the system sizes of TM data are 
larger than for ELS data".} I do not know about "currently" and "standard" 
but only one result is cited in \cite{2} for systems $\sim 50^3$ \cite{5}. 
With my procedure used, authors of \cite{2} obtained  this result with 
uncertanty $\nu=1.25\div 1.75$, which is not drastically different from 
uncertainty $0.8\div 1.7$ that I have derived from Zharekeshev and Kramer 
data \cite{6}.  So, it is admitted that the largest scale data give 
uncertainty $1.25\div 1.75$  and not 1\% \cite{4}.  At the present stage, I 
am quite satisfied with such progress.

\vspace{5mm}

2. Authors of \cite{2} claim, that my procedure is not interesting 
because it does not give increase of accuracy. They are mistaken here:
I had not purpose to increase accuracy. I wanted to dig out the real 
uncertainty of results. According to \cite{2}, estimation of error
given by my procedure approximately corresponds to a scattering of 
results due to different authors. So this estimation is reliable and 
reasonably conservative. I can hardly dream of this. By the way, only "people 
with experience in FSS" can believe in accuracy of several percents,
looking at Fig.~1 of \cite{2}.

In fact, authors of \cite{2} have overlooked the essense of my 
suggestions. I have introduced some function $f(L)$, in order to have a 
convinient languige for discussion of scaling corrections. My 
suggestions are simple. Show your function $f(L)$, so that we can
have a look at it. Then it will be evident, has this function so good 
power-law behavior as you say, or not. It will be clear,
is there change of the effective $\nu$ with the length scale, or not.
Then we can compare such functions for  different quantities and  
different models, in order to estimate a systematic error. Then we can see, 
how this error change with $L$ and so on.

I do not insist, that my procedure for determination of the function $f(L)$ 
is the best. In fact, it is useful in the case of rather poor raw data.
If these data are sufficiently detailed and accurate, it is more preferable 
to extract the function $f(L)$ by direct linearization (in fact, it was 
admitted in \cite{1}).  Such procedure was used in a number of papers and I 
have no priority in it.  But it was only recently, that explicit plots of 
such functions have appeared \cite{7,8}.
\vspace{5mm}

3. I agree with \cite{2a}, that in the general case a fixed point depends on 
$L$ and irrelevant parameters are uncritical. 
The fixed point $\mu^*$ has a regular dependence in $a/L$ and the main 
correction is the ordinary surface effect. With this dependence 
taken into account, my Eq.(5) takes a form
$$
Q(\mu)=F\left\{\mu^*(L/a), A_1(\tau) s^{y_1},\, A_2(\tau) s^{y_2},\,
     A_3(\tau) s^{y_3},\,\ldots  \right\} \,\,,
\eqno(5')
$$
Putting $A_1(\tau)=b_1\tau$, $A_i(\tau)=c_i+b_i\tau$ ($i\ge 2$), we have 
instead (6)
$$
Q(\tau,L)=F\left\{a/L,\,b_1\tau (L/a)^{y_1},\,(c_2+b_2\tau)(L/a)^{y_2},\,
     (c_3+b_3\tau) (L/a)^{y_3},\,\ldots  \right\}  \,\,,
\eqno(6')
$$
For $L\gg a$ we have instead (7) and (8) correspondingly
$$
Q(\tau,L)=\left[ F\left\{0,0, 0,0,\ldots  \right\} + B_0(a/L) +B_2 
(L/a)^{y_2}+B_3 (L/a)^{y_3},\,\ldots  \right]+
$$
$$
+\tau \left[ C_1 (L/a)^{y_1} +C_2 (L/a)^{y_2}+C_3 (L/a)^{y_3},\,\ldots  
\right]
\equiv F_0(L/a) + \tau f\left(L/a \right)\,,
\qquad  \tau (L/a)^{y_1} \ll 1
\eqno(7')
$$
and
$$
Q(\tau,L)\approx F\left\{0,b_1\tau (L/a)^{y_1},\,0,\,0,\,\ldots  \right\}
\equiv G\left\{\tau (L/a)^{y_1} \right\} \,\,,
\qquad  \tau (L/a)^{y_1} \agt 1 \,\,.
\eqno(8')
$$
Correction to the critical point
$$
\tilde\tau=\tau +\Delta(a/L)\,\,,\qquad
\Delta(a/L)=\left[F_0(L/a) - F_0(0)\right]/f(L/a) 
$$ 
is essential for small $\tau$ in Eq.($7'$)  but is irrelevant in Eq.($8'$).  
So unification ($7'$) and ($8'$) has a form:  
$$ 
Q(\tau,L)= G\left\{\tilde\tau f(L/a) \right\} \,\,.  
\eqno(9')
$$
One should shift the curves for different $L$,  
so as they have a common intersection point, and use the same procedure as  
in \cite{1}. Such shift is routinely used in the most of papers. There
was no need in it for the data of \cite{6}.

For $d=4-\epsilon$ the first irrelevant exponent $y_2$ is of the order
of $\epsilon$ \cite{8a} and linearization in the parameter $c_2(L/a)^{y_2}$ 
is invalid for $L\alt L_0\sim a\exp\{const/\epsilon\}$. The corresponding 
dependence on $L$ is inessential in the restricted intervals of $L$ but
there is a slow drift of results on the large scale $L_0$. May be,
it is a clue to the whole problem.

\vspace{5mm}

4. There are a lot of vague speculations in \cite{2a} concerning 
a number of nonlinear parameters. Authors were able to calculate
that in my procedure this number is even greater than in their one.
In spite of these speculations, it it evident to anyone that

\noindent
\quad (a) all steps of my procedure are well defined and unambiguous.

\noindent
\quad (b) authors of \cite{2a} did not give a clear answer to
a clear question: how do they avoid ambiguity of their procedure?

With modifications of \cite{2a}, my procedure becomes ambiguious too.
There is no sense to discuss the results of the unclear procedure.

\vspace{5mm}

5. One can carry out a very simple experiment. Take some function, f.e.
$f(x)=x^2/(1+x)$, add some noise to it, and try to find its asymptotic 
behavior $A x^\alpha$ at $x\to\infty$, using only values of $f(x)$ for 
$x=1,2,\dots,20$. You will find it rather difficult to formulate 
a procedure that gives the reliable and reasonably conservative estimation of 
errors. The more or less successful procedure is as follows. Take an 
interval $(x_{min},20)$ and increase $x_{min}$ until a simplest fit
$\log f(x)=C+\alpha\log x$ give for $\chi^2$  its 
normal value $\approx n$, where $n$ is a number of points\,\footnote{\,
We suppose that this number is much greater than a number of parameters.
}. It is 
a necessary condition for a systematic error to be of the order of 
a statistical one. Now increase $x_{min}$ {\it a little more}: then the 
formal statistical error will give more or less reasonable estimation of the 
real error.  {\it A little more} is rather subjective and depends on the 
degree of conservatism. This procedure is successful, if a noise is 
uncorrelated and you have reliable estimation of its amplitude. In the 
general case such procedure is minimal and gives only the lower bound of the 
error.  In my experience, related with a paper \cite{9} and its development 
\cite{10}, the noise was strongly correlated and a formal statistical error 
was typically two orders less than a real one (even for the good values of 
$\chi^2$). I can only wonder that some people are so confident in their 
accuracy.

The given example is simple, because corrections to asymptotics  {\it are
regular}. You can increase accuracy of $\alpha$ by many orders, fitting $\log 
f(x)$ as $C_0+\alpha \log x +C_1/x+C_2/x^2+\ldots$ (this fit is linear and, 
consequently, unique). It is not the case that is 
interesting for us.

If a structure of corrections is power-like but unknown, you should use a 
nonlinear fit $f(x)=Ax^\alpha+A_1 x^{\alpha_1}+A_2 x^{\alpha_2}+\ldots$
and you immediately meet with a problem of many minima.  Their origin is 
rather simple. Take an arbitrary succession $\alpha,\alpha_1,\alpha_2\ldots$
(such that $\alpha>\alpha_1>\alpha_2>\ldots$) and use for $f(x)$ a linear fit
$Ax^\alpha+A_1 x^{\alpha_1}+A_2 x^{\alpha_2}+\ldots$  (only 
$A,A_1,A_2,\ldots$ are changed). It will be successful with sufficient number 
of terms and $\chi^2$ takes its normal value. Take this result as "a zero
approximation" and use the general nonlinear fit (not only $A,A_1,A_2,\ldots$
but also $\alpha,\alpha_1,\alpha_2,\ldots$ are changed). The quantity $\chi^2$
cannot be essentially lower than its normal value, and can change only 
slightly. So, there is a local minimum for it, close to "a zero 
approximation". Taking different successions 
$\alpha,\alpha_1,\alpha_2\ldots$,
you can generate a great number of such minima. A value of $\alpha$ is 
different in different successions. So you can obtain an {\it arbitrary}
result for it\,\footnote{\,It is restricted only by the assumptions made.} by 
such procedure.

I think, it is clear now, why one should not believe in any results
obtained with such fits.  If a structure of corrections is unknown, one 
hardly can invent something better than the simplest procedure given in 
beginning of this section.

\vspace{5mm}

6. After \cite{1} I had a constructive discussion with 
P.~Marko{$\rm\check  s$}\,\footnote{ \,He has no responsibility for my 
conclusions.} and now I have better understanding of a general situation. 
Here are my impressions of it.

\vspace{2mm}

\quad (i) There is a good paper \cite{8}, where the transfer-matrix method and
the system 
sizes up to $L=24$ were used. The raw data are rather detailed and
accurate, so a function $f(L)$ can be obtained with good 
accuracy\,\footnote{\, I think,  this accuracy is a little 
exaggerated. One can see in Fig.~2 of \cite{8} that a number of points
deviates from the general smooth dependence in a quantity that is not
controlled by the given error.} by 
direct linearization of $W$-dependence.  More than that, Fig.~2 of this paper 
shows a number of such functions (in the log--log coordinates) for the 
different Lyapunov exponents.  Some observations can be made in this figure.
 For $L\alt 10$ the slopes of the curves differ very strongly and correspond
to values of $\nu$ in the interval $0.8\div 1.5$.  For $L\agt 10$ practically 
all curves change their slopes and the more or less unique slope (with 
$\nu\approx 1.5$) arises. Nevertheless, the curve for the least exponent 
$z_1$ (which, according to \cite{8}, should reach its asymptotics most 
quickly) has a tendency to change its slope in the opposite direction. 
For $L>12$, this 
slope is compatible with a value $\nu=1.25$, if a point for $L=24$ (evident 
outliar) is excluded.  One can see, that a possible change of the effective 
$\nu$ with $L$ is not a fantasy.

\vspace{2mm}

\quad (ii) There is a lot of high precision results for $L\alt 15$ and they indeed 
give  a value $\nu\approx 1.5$. Let us assume, that this value 
is not true, but effective. Then it should change with increase of $L$. In 
practice, one can double \cite{6,8} or triple \cite{5} this size, though 
there are not many such attempts. With such increase of $L$, a situation 
cannot improve drastically. One can be able only to see some change of $\nu$ 
with $L$. There is a number of reasons, why this change could not be observed 
reliably:

\noindent
\quad (a) Accuracy of the raw date is worse by an order of magnitude
for $L\agt 15$.

\noindent
\quad (b) A need for a careful treatment was underestimated. The researches
were satisfied that the large scale data roughly correspond to 
well-established results for small $L$, though the most of fits
are not satisfactory.

\noindent
\quad (c) When linearization of $W$-dependence was used, the interval of 
linearization was kept fixed. It gives the wrong tendency due to 
increase of nonlinear effects for the large $L$ (see footnote 5 in \cite{1}). 
This tendency is not large in magnitude, but the whole situation is rather 
unstable. One can see in Fig.~2 of \cite{8} that a shift of the large $L$ 
data in one standard deviation produces essential change of results.

As a consequence, the large $L$ data appeared to be {\it practically 
useless}.  In fact, this conclusion is supported by authors of \cite{2}, who 
refer primarily to results for small and not for large $L$.

\vspace{2mm}

\quad (iii) If the raw data have a quality comparable with \cite{8},
 the adequate procedure should contain, in our opinion, the following
stages:

\noindent
\quad (a) In order to have
the efficient use of the $\chi^2$ procedure, the statisical uncertainties
of the raw data are not simply estimated, but are calculated independently 
for each point in the course of some formal procedure. Probably, it can be 
done with the use of different  realizations of a random potential.

\noindent
\quad (b) The function $f(L)$ is obtained by linearization of $W$-dependence.
The interval of linearization is adjusted in accordance with $\chi^2$ and the 
error is properly estimated. It should be done independently for each $L$, 
but with the use of the same routine.  More conservatism is desirable at 
this stage.

\noindent
\quad (c) The function $f(L)$ is plotted. It is fitted by the power law dependence 
in the interval $(L_{min},L_{max})$ where $L_{min}$ is adjusted with the use 
of $\chi^2$ and $L_{max}$ is the largest $L$ that is possible to reach.  
Minimization of $\chi^2$ is a clever procedure and it provides a necessary 
compromise between the high accuracy data for small $L$ and the poor accuracy 
data for large $L$: the latter are properly weighted, but all useful 
information is extracted of them. Adjustment of $L_{min}$ allows to exclude 
the main body of a systematic error,  which is contained in the small $L$ 
data. Excess of conservatism at the stage (b) is partially compensated, 
because smaller $L_{min}$ becomes admittable.

\noindent
\quad (d) The rest of a systematic error can be controlled by comparison
of results for different quantities and different models.

I have impression, that nobody had used this procedure in a full 
extent\,\footnote{\,Sensitibility of results to restriction
of the intervals in $W$ and $L$ was observed by MacKinnon\cite{12}.}. 

\vspace{2mm}

\quad (iv) Some comments concerning the 4D case. The most advanced results
are those of Zharekeshev and Kramer \cite{7}. With $\nu=1$, they have a 
parameter $\tau(L/a)^{1/\nu}$ (in designations of \cite{1}) of the order 
unity, and a condition for linearization of $W$-dependence is marginally 
fulfilled. With $\nu=1/2$ (as it should be), the above parameter is $\sim 10$ 
and they are deeply in the nonlinear regime.  So, a possibility $\nu=1/2$, 
in fact, was not tested numerically.

Some confirmation of these arguments can be found in \cite{13}.
Fig.~1,a  of this paper shows $W$-dependences for another quantity
(Lyapunov exponent, but not level statisics) which are essentially
more detailed in comparison with \cite{7}. Indeed, they are 
strikingly nonlinear. 
More attention should be given to investigation
of these nonlinearities.

\vspace{6mm}

This work was financially supported by  INTAS (Grant 99-1070) and 
RFBR (00-02-17129).

\end{document}